\documentclass[letter]{aa}

\usepackage{graphicx}
\usepackage{natbib}
\bibpunct{(}{)}{;}{a}{}{,}
\usepackage{txfonts}
\def\deg{$^\circ$}
\def\arcsec{"}

\def\aap{A\&A}

\def\apj{ApJ}
\def\apjl{ApJL}

\def\mnras{MNRAS}

\def\kms{km~s$^{-1}$}

\def\lesssim{\mathrel{\hbox{\rlap{\hbox{\lower4pt\hbox{$\sim$}}}\hbox{$<$}}}}
\def\gtrsim{\mathrel{\hbox{\rlap{\hbox{\lower4pt\hbox{$\sim$}}}\hbox{$>$}}}}
\begin{document}
\title{The 2011 outburst of the recurrent nova \object{T\,Pyx}.\\ Evidence for a face-on bipolar ejection.}

\author{O. Chesneau\inst{1} 
          \and
          A. Meilland\inst{1}
          \and
          D.~P.~K. Banerjee\inst{2}
          \and
          J.-B. Le Bouquin\inst{3}
          \and 
          H. McAlister\inst{4,5}
          \and
          F. Millour\inst{1}
          \and
          S.T. Ridgway\inst{6}
          \and
          A. Spang\inst{1}
       \and
       T. ten~Brummelaar\inst{5}
       \and
          M. Wittkowski\inst{7}
          \and 
          N.M. Ashok\inst{2}
          \and
          M. Benisty\inst{8}
          \and
          J.-P. Berger\inst{9}
          \and
          T. Boyajian\inst{4}
          \and
          Ch. Farrington\inst{5}
          \and
          P.J. Goldfinger\inst{5}
           \and
          A. Merand\inst{9}
          \and 
          N. Nardetto\inst{1}
          \and
          R. Petrov\inst{1} 
          \and
          Th. Rivinius\inst{9}
           \and
           G. Schaefer\inst{4}
           \and
           Y. Touhami\inst{4}      
           \and
           G.~Zins\inst{3}
           \fnmsep \thanks{Based on observations made with CHARA at Mount Wilson observatory and the VLTI at Paranal Observatory under program 287.D-5012, 287.D-5023, 087.C-0702}
	}
	   \offprints{Olivier.Chesneau@oca.eu}

\institute{
UMR 6525 Fizeau, Univ. Nice Sophia Antipolis, CNRS, Obs. de la C\^{o}te d'Azur, 
Bvd de l'Obs., BP4229 F-06304 NICE Cedex 4
              \and
Physical Research Laboratory, Navrangpura, Ahmedabad, Gujarat, India
\and
UJF-Grenoble 1/CNRS-INSU, Institut de Plan\'etologie et d'Astrophysique de Grenoble (IPAG), UMR 5274, Grenoble, France
\and
Georgia State University, P.O. Box 3969, Atlanta GA 30302-3969, USA
 \and
CHARA Array, Mount Wilson Observatory, 91023 Mount Wilson CA, USA
\and
National Optical Astronomy Observatories, 950 North Cherry Avenue, Tucson, AZ, 85719, USA
\and
European Southern Observatory, Karl-Schwarzschild-Strasse 2, D-85748 Garching bei M\"unchen, Germany
\and
Max Planck Institut f\"ur Astronomie, K\"onigstuhl 17, 69117 Heidelberg, Germany 
\and
European Southern Observatory, Casilla 19001, Santiago 19, Chile
}
\date{Received, accepted.}
\abstract
{}
%Aims
{T\,Pyx is the first recurrent nova historically studied, seen in outburst six times between 1890 and 1966 and then not for 45 years. We report on near-IR interferometric observations of the recent outburst of 2011. We compare expansion of the H and K band continua and the Br$\gamma$ emission line, and infer information on the kinematics and morphology of the early ejecta. }
%Methods
{We obtained near-IR observations of \object{T\,Pyx} at dates ranging from t=2.37d to t=48.2d after the outburst, with the CLASSIC recombiner, located at the CHARA array, and with the PIONIER and AMBER recombiners, located at the VLTI array. These data are supplemented with near-IR photometry and spectra obtained at Mount Abu, India.} 
%Results
{Slow expansion velocities were measured ($\leq$300\kms) before t=20d. From t=28d on, the AMBER and PIONIER continuum visibilities (K and H band, respectively) are best simulated with a two component model consisting of an unresolved source plus an extended source whose expansion velocity onto the sky plane is lower than $\sim$700\kms. The expansion of the Br$\gamma$ line forming region, as inferred at t=28d and t=35d is slightly larger, implying velocities in the range 500-800\kms, still strikingly lower than the velocities of 1300-1600\kms inferred from the Doppler width of the line. Moreover, a remarkable pattern was observed in the Br$\gamma$ differential phases. A semi-quantitative model using a bipolar flow with a contrast of 2 between the pole and equator velocities, an inclination of i=15\deg\ and a position angle P.A.=110\deg\ provides a good match to the AMBER observables (spectra, differential visibilities and phases). At t=48d, a PIONIER dataset confirms the two component nature of the H band emission, consisting of an unresolved stellar source and an extended region whose appearance is circular and symmetric within error bars.}
%Conclusions
{These observations are most simply interpreted within the frame of a bipolar model, oriented nearly face-on. This finding has profound implications for the interpretation of past, current and future observations of the expanding nebula.}
\keywords{Techniques: high angular
                resolution; (Stars:) novae, cataclysmic variables; individual: \object{T\,Pyx};
                Stars: circumstellar matter}

\titlerunning{The 2011 outburst of the recurrent nova \object{T\,Pyx}.}
\authorrunning{Chesneau et al.}

\maketitle

\section{Introduction}
A classical nova eruption results from a thermonuclear runaway on the surface of a white dwarf which is accreting material from a companion star in a close binary system. T\,Pyxidis (\object{T\,Pyx}) is a unique recurrent nova that was in outburst six times between 1890 and 1966 (intervals of $\sim$20yr). \object{T\,Pyx} was discovered in outburst at a visual magnitude of 13.0 on 2011 April 14.29 UT (JD=2455665.79); which we take as t$_0$=0  \citep{2011CBET.2700....1W}. This is the first outburst of \object{T\,Pyx} since December 7, 1966, nearly 45 years before.

The evolution of the nova is relatively slow, thereby providing
time and scope for organizing joint observations with optical interferometry arrays such as CHARA and the VLTI.
\object{T\,Pyx} is surrounded by an interesting nebula in expansion that has been investigated by the HST during more than 10yr \citep[and references therein]{2010ApJ...708..381S}. The knots are expanding in the plane of the sky with velocities ranging from roughly 500 to 715 \kms.  In contrast, the velocities inferred from Doppler widths of the ejecta of recent outbursts were observed to be much faster at about 1500 \kms. Although \object{T\,Pyx} is a well-observed system, it still has many mysteries. Why did the ejecta expand so slowly in the plane of the sky? An important spectroscopic study of the binary system from \citet{2010MNRAS.409..237U} provided evidence of a low-inclination for the system orbit (i=10$\pm$2\deg), a particularly important constraint for the interpretation of interferometric data, as it appears that the ejecta emitted around these outbursting sources are rarely spherical. 
%High resolution optical spectroscopy obtained during pre-maximum stage \citep{2011ATel.3376....1S} show Balmer series and He I/II with P %Cyg components. Then the P Cyg absorption became more evident around May 6.9, and spectra obtained around exhibit lines with very broad %emission component at +2000 to +3000 km/s (Banerjee et al. in preparation?). 

This letter presents optical interferometry measurements obtained from different facilities which provide important information when included within a common frame of interpretation. The observations are presented in
Sect. \ref{Observations}. In Sect. \ref{Analysis} we analyze the continuum measurements by means of simple geometrical model, and the differential observables through the Br$\gamma$ line using a simple model and then discuss the results
in Sect. \ref{Discussion}.
%%%%%%%%%%%%%%%%%%%%%%%%%%%%%%%%%%%%%%%%%%%%%%%%%%%%%%%%%%%%%%%%%%%%%%%%%%%%%%%%
%%%%%%%%%%%%%%%%%%%%%%%%%%%%%%%%%%%%%%%%%%%%%%%%%%%%%%%%%%%%%%%%%%%%%%%%%%%%%%%%%%%%%%%%%%%%%%%%%
%JOURNAL DES OBSERVATIONS
%%%%%%%%%%%%%%%%%%%%%%%%%%%
%\onltab{1}{
\begin{table*}
\caption[]{Journal of interferometric observations.}
\centering
\begin{tabular}{c c c c c c c c c}
	\hline
	\hline
	date & MJD & t-t$_0$ & Instrument & Base&	\multicolumn{2}{c}{projected baselines}& Calibrators$^{\mathrm{a}}$\\
			 &2450000.5+&Phase & &&	length [$\mathrm{m}$]&	P.A. [\deg]& \\
	\hline
	2011/04/17& 5668.16 & 2.92 & CLASSIC & W1-W2& 107.4 &$-$87 & HD\,78752, HD\,79290\\
	2011/04/23& 5674.04 & 8.81 & AMBER & K0-A1-I1 & 45.4/100.3/120.2&  $-$159.0/96.2/ $-$105.0& HD\,73947\\ 
	2011/04/26 & 5678.06 & 12.81 & PIONIER& A1-G1-I1-K0& 74/94/42/113/82/44&  $-$52/$-$74/$-$127/81/39/23& HD\,78739\\ 
	2011/04/28& 5679.18 & 13.93 & CLASSIC & W1-E2& 202.1/213.3&  $-$79.4/82.8& HD\,78752, HD\,79290\\
  2011/05/12& 5694.00 & 28.76 & AMBER & UT1-3-4& 59.5/95.6/117.7&  127.1/43.1/73.3& HD\,73947\\
	2011/05/20& 5701.02 & 35.77 & AMBER & UT1-3-4& 56.2/89.9/105.6&  136.9/$-$133/78.6& HD\,73947, HD\,87303\\
	2011/06/01& 5713.99 & 48.74 & PIONIER & D0-G1-H0-I1 & 68/47/63/67/37/40 & 171/96/132/25/57/180 & HD\,78739 \\
	\hline
\end{tabular}
	\label{obslog}
	\begin{list}{}{}
%	\item[$^{\mathrm{a}}$]  t0=2455665.79. 
	\item[$^{\mathrm{a}}$] Calibrator angular diameters from SearchCal@JMMC\citep{2006A&A...456..789B}: HD\,78752 (G0V, 0.22$\pm$0.02mas), HD\,79290 (A0V, 0.13$\pm$0.01mas), HD\,73947 (K2III, 0.86$\pm$0.02mas), HD\,87303 (K2III, 0.90$\pm$0.07mas), HD\,78739 (K0III, 0.32$\pm$0.02mas).
	\end{list}
\end{table*}
%}
\section{Observations}
\label{Observations}

\begin{figure}
 \centering
\includegraphics[width=7.9cm]{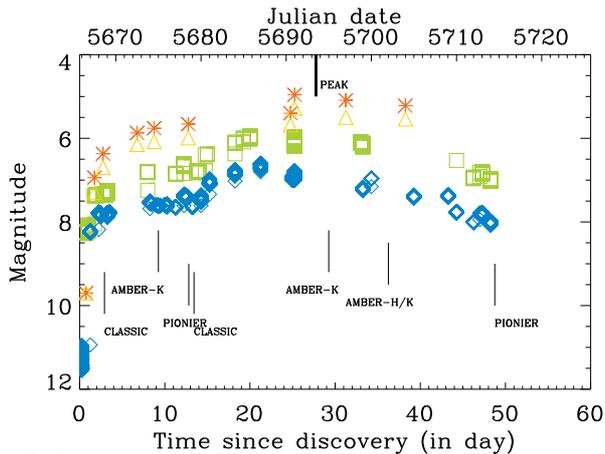} 
 \caption[]{Light curve of \object{T\,Pyx} with the dates of the optical interferometry observations. Blue diamonds indicate a subset of AAVSO data in V, and green squares in I. Orange triangles and red stars indicate H and K band photometry from Mt Abu (India) \label{fig:ABU}. \label{fig:lightCurve}}
\end{figure}

{\it Near-infrared JHK photometric and spectroscopic observations} were
obtained on a regular basis from the 1.2m telescope at the Mt. Abu
Observatory, India.  These measurements helped to prepare the
interferometric observations and to evaluate the relative contribution of
the various continuum and line components (Fig.\ref{fig:ABU}).  Initial observations
are  reported  in \citet{2011ATel.3297....1B} while a fuller study is in
preparation.

%\onlfig{2}{
\begin{figure}
 \centering
 \includegraphics[width=7.9cm]{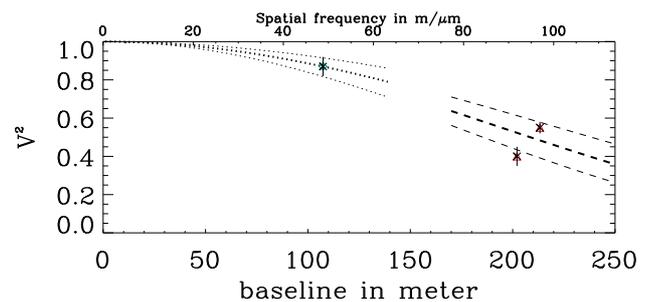}
 \caption{K-band interferometric visibilities obtained with CLASSIC at t=2.92 (green diamond) and t=13.93 (red triangles). The thick dotted and dashed lines indicate are the UD curves corresponding to Table 2.   \label{fig:CLASSIC}}
\end{figure}
%} 
 
Prompt {\it broad-band interferometric observations} were secured with CLASSIC, a two-telescope high sensitivity system located at CHARA on Mt. Wilson \citep{2005ApJ...628..453T}. Despite the faintness and low declination of the source, observations in the K-band were obtained at t=2.92d (K$\approx$6.4, from Mt Abu observations) and t=13.93d (K$\approx$5.7). The log of the observations is presented in Table\,\ref{obslog} and the data in Fig.\ref{fig:CLASSIC}.

Several {\it interferometric observations at medium spectral resolution} (R=1500) across the Br-$\gamma$ line were obtained with AMBER, a 3-telescope combiner located at the VLTI  \citep{2007A&A...464....1P}. The first observations were performed with the 1.8m Auxiliary Telescopes (ATs) at t=8.81d, when the source was below the interferometric sensitivity limit of AMBER (K$\approx$5.7), but a useful spectrum was obtained. The second and third measurements, obtained with the 8.2m Unit Telescopes (UTs) at t=28.76 (K$\approx$4.9) and t=35.77d (K$\approx$5), provided good quality dispersed visibilities, closure and differential phases (see Fig.\ref{fig:AMBER}). Unfortunately, the calibrator measurement for the last date is of poor quality preventing any reliable calibration of the absolute visibility.

\begin{figure}
 \centering
  \includegraphics[width=6.9cm]{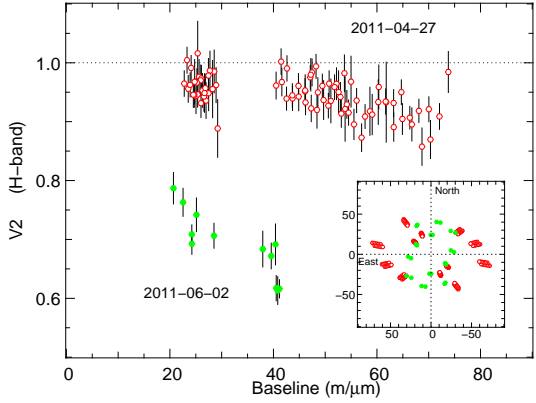}
 \caption{H-band interferometric visibilities obtained with PIONIER at t=12.81  (open red) and t=48.74 (filled green). The corresponding uv-plane is displayed in the subpanel. \label{fig:PIONIER}}
\end{figure}

{\it Imaging broad-band interferometric observations} were obtained at t=12.81d (H$\approx$6) with the PIONIER visitor instrument \citep{2010SPIE.7734E..99B,2011AA.PIONIER}. These observations provided the simultaneous measurement of 6 absolutely calibrated visibilities and 4 closure-phases in the H-band, therefore allowing the study of the spatial morphology of the near-infrared emission. A critical second observation was obtained at t=48.74d (H$\approx$6), again with the ATs (Fig.\ref{fig:PIONIER}).
 
%%%%%%%%%%%%%%%%%%%%%%%%%%%%%%%%%%%%%%%%%%%%

\section{Analysis}
\label{Analysis}
\begin{table*}
\caption[]{Analysis of the V$^2$ using geometrical models.\label{models}}
\centering
\begin{tabular}{c c c  |c | c c c }
	\hline
	\hline
Instrument& Spectral Band & t-t$_0$ & Single component model &\multicolumn{3}{c}{Double components model}\\
%	\hline
 &		&	day & UD diameter  &Unres.  Flux  &	 UD Flux &	UD diam.   \\
&		&	 & [$\mathrm{mas}$] & [$\mathrm{\%}$] &	 [$\mathrm{\%}$] &	[$\mathrm{mas}$] \\
	\hline
CHARA/CLASSIC & broad K & 2.92& 1.$\pm$0.2& - & -  & -\\
VLTI/PIONIER & broad H & 12.81& 0.6$\pm$0.1 &  -&  -& -\\
CHARA/CLASSIC & broad K & 13.93 & 1.12$\pm$0.14 &  -& - & -\\
VLTI/AMBER & 2.1$\pm$0.05$\mu$m& 28.76 & 2.58$\pm$0.3& 65$\pm$12 & 35$\pm$8 & 7.3$\pm$0.3\\
VLTI/PIONIER & broad H & 48.74 & 2.23$\pm$0.1& 83$\pm$9& 17$\pm$2 & 8.5$\pm$0.2 \\
	\hline
\end{tabular}
\end{table*}

The absolute visibility measurements were fitted with simple geometrical models using the LITpro software \citep[JMMC]{2008SPIE.7013E..44T}.
The results are shown in Table\,\ref{models}. A simple uniform disk (UD) model, i.e. a circular disk of uniform brightness in the plane of the sky, was fitted to the measurements for the early observations. For later observations, a two component model consisting in an unresolved component, and a co-centered uniform disk was used. No evidence of asymmetry was detected in the data.
 
\begin{figure}
 \centering
 \includegraphics[width=7.9cm]{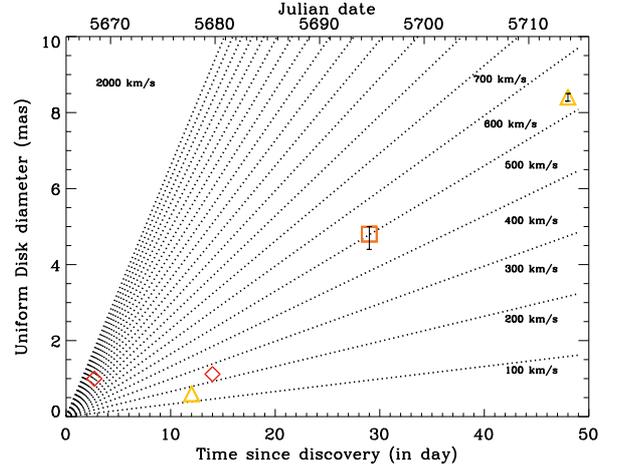}
 \caption[]
{Result of the Uniform Disk (UD) estimates from the continuum $V^2$ measurements from the various interferometers. Before t=20d, the source size is estimated using a single UD in the H and K bands (PIONIER: orange triangle; CLASSIC: red diamonds). The last points indicate the extended source in the double component model (AMBER: red square; PIONIER: orange triangle) \label{fig:fit}}
\end{figure}

The first K-band CLASSIC measurement, obtained with a $\sim$100m baseline was consistent with a weakly resolved source, while the second set (t=13.93d) with a baseline about twice as long, shows a resolved source whose size does not seem to be dramatically changed. The first PIONIER measurements (t=12.81d) provided a H-band UD diameter significantly smaller than the CLASSIC K band measurements, even taking into account the small time difference between the two measurements. This effect cannot be attributed to emission lines seen at these dates, that contribute less than 10-15\% of the flux in the H band and less than 5\% in the K band. A large scale component with a rising flux contribution in the K band may account for the observations, and is consistent with a H-K flux difference of 0.33 mag that is observed consistently during the event. Assuming a two component model, with a fixed contribution of 70$\pm$10\% from a 'stellar' source with a diameter of 0.6mas, the extended K-band source should have a diameter larger than 1.5-2mas, and therefore be almost fully resolved by CLASSIC. Assuming a distance of D=3.5$\pm$1kpc from \citet{2010ApJ...708..381S}, that may be a lower limit \citep{2011A&A...533L...8S}, the expansion velocity inferred from such an extended component is about 500\kms, while the H-band core expansion is estimated to be $\sim$100\kms (Fig.\ref{fig:fit}). Interestingly, the FWHM of the Br$\gamma$ measured from the AMBER spectrum at t=8.81d is 590\kms, i.e. is consistent with our hypothesis of the expanding extended  component (Fig.\ref{fig:AMBER}). 

Then follows a second epoch during which AMBER interferometric data were obtained on two dates, yet providing calibrated visibilities only at t=28.76d. A single UD does not account well for the observed visibilities at t=28.76d (reduced $\chi_r^2$=35), the two component model provides a better fit to the data ($\chi_r^2$=5).  This gives an upper limit for the expansion of the extended  continuum component of 700\kms. This is in contrast with the width of the  Br$\gamma$ line, for which we measured a FWHM of 1050$\pm$50\kms. Furthermore, the Doppler velocity associated with the P-Cygni absorption in the line is found to be $-1450\pm$100\kms (Fig.\ref{fig:AMBER}). The Br$\gamma$ line-forming region should therefore expand much faster, and a large visibility drop should be measured through the Br$\gamma$ line, as seen for \object{RS\,Oph} \citep{2007A&A...464..119C}. However, the dispersed Br$\gamma$ visibilities are only slightly lower than the nearby continuum, implying a moderate diameter increase of less than 10\%. One week later, the visibilities dropped in the line indicating a large expansion of the Br$\gamma$ line-forming region (taking into account the 25\% increase of the line flux). At that time the line FWHM was measured to be 1600$\pm$50\kms\ and the P-Cygni absorption indicated a wind velocity of 1800$\pm$100\kms. The differential phases show a complex structured signal that can be described by 2 opposite S-shaped signals, with variations related to the baseline lengths and P.A. dependency. The pattern is symmetrical about the line center, with the width and amplitude of the signal increasing between the two dates, although the amplitude never went beyond 10\deg\ (again by contrast with what was observed for \object{RS\,Oph}). 

The last observations, performed at t=48.74d with PIONIER, bring complementary and crucial information, owing to the larger $uv$ coverage involved. The H band source is now well resolved and departs from a simple model. A good fit ($\chi^2=1.1$) is reached using the two component model (Table\,\ref{models}). The extended component estimated expansion velocity in the plane of the sky is lower than 700\kms. The closure phases do not exceed 2.5\deg. Moreover using a flattened structure for the extended component does not improve the fit and constrains the aspect  ratio to 1$\pm$0.07. This implies that the complex yet weak phase signal seen by AMBER originates from a source with a predominantly symmetrical appearance.

\section{A face-on bipolar event}
\label{Discussion}

\begin{figure*}
 \centering
\includegraphics[width=7.5cm]{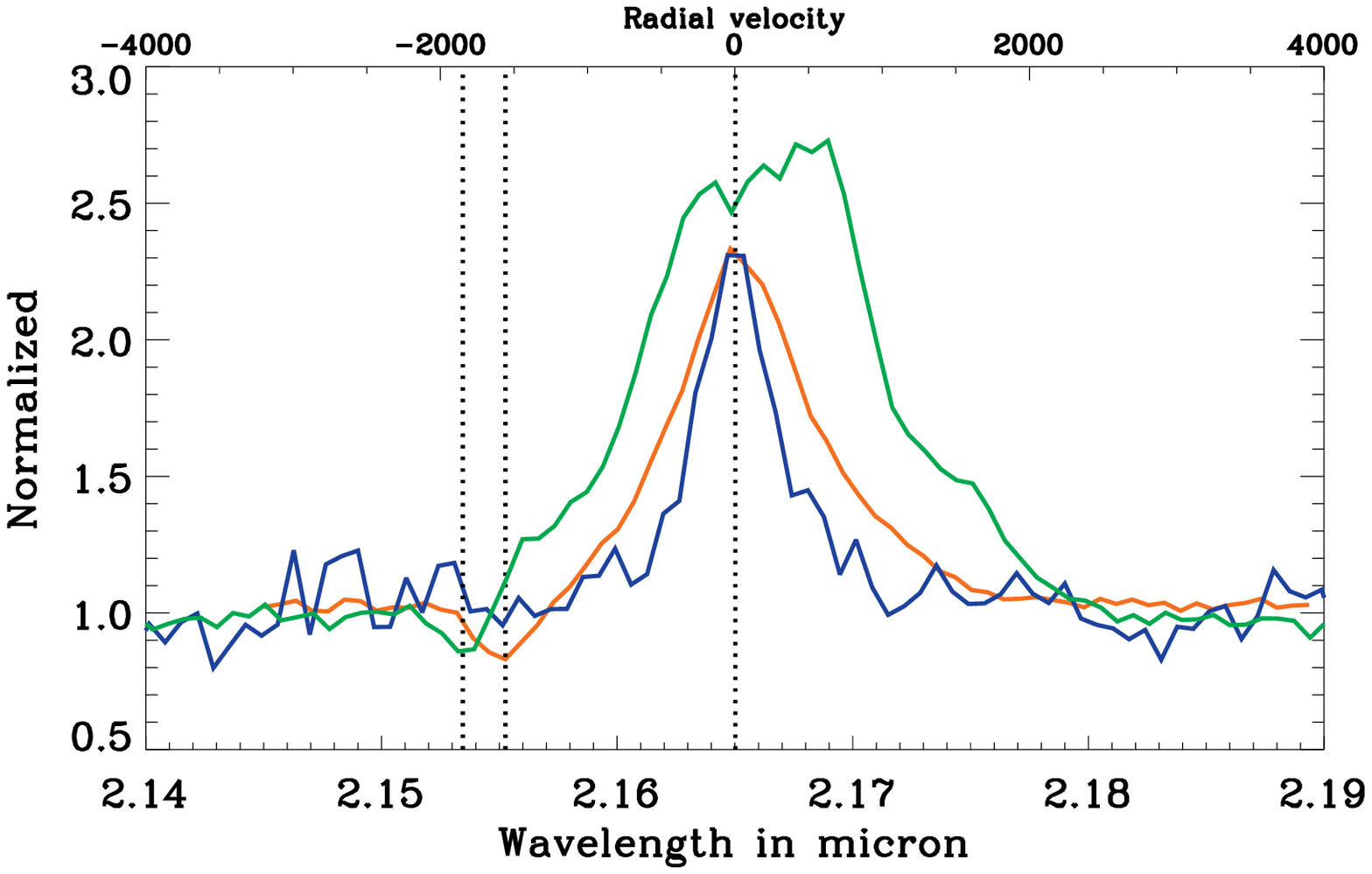}
\includegraphics[width=15.2cm]{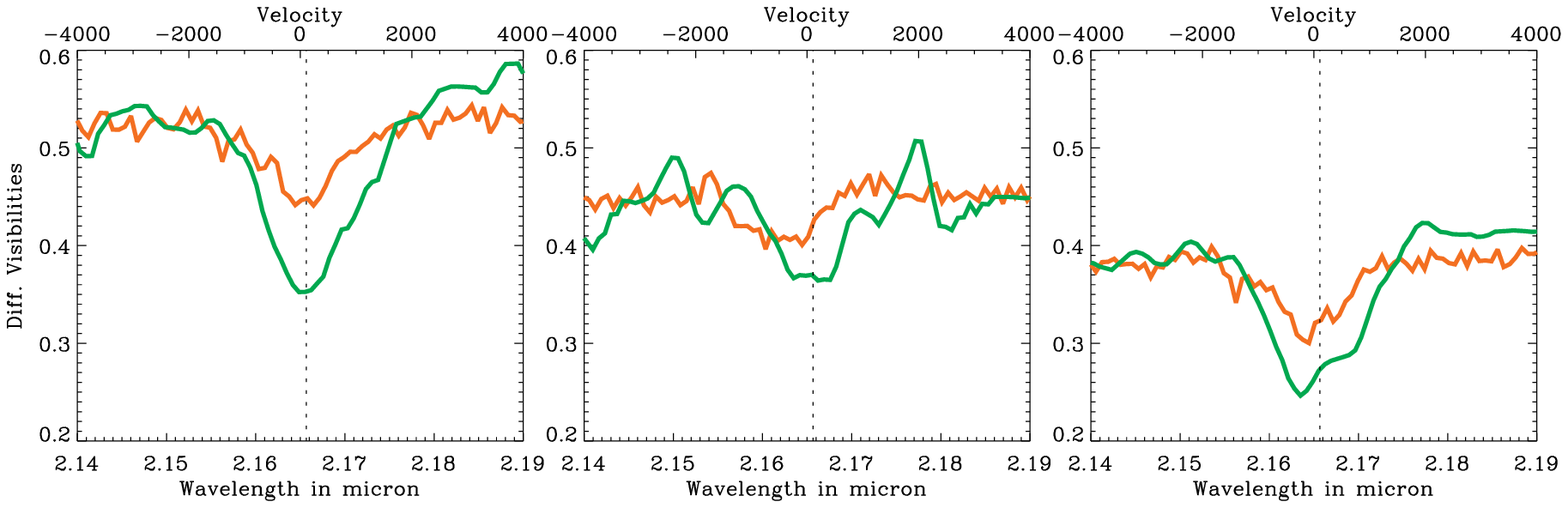}
\includegraphics[width=15.2cm]{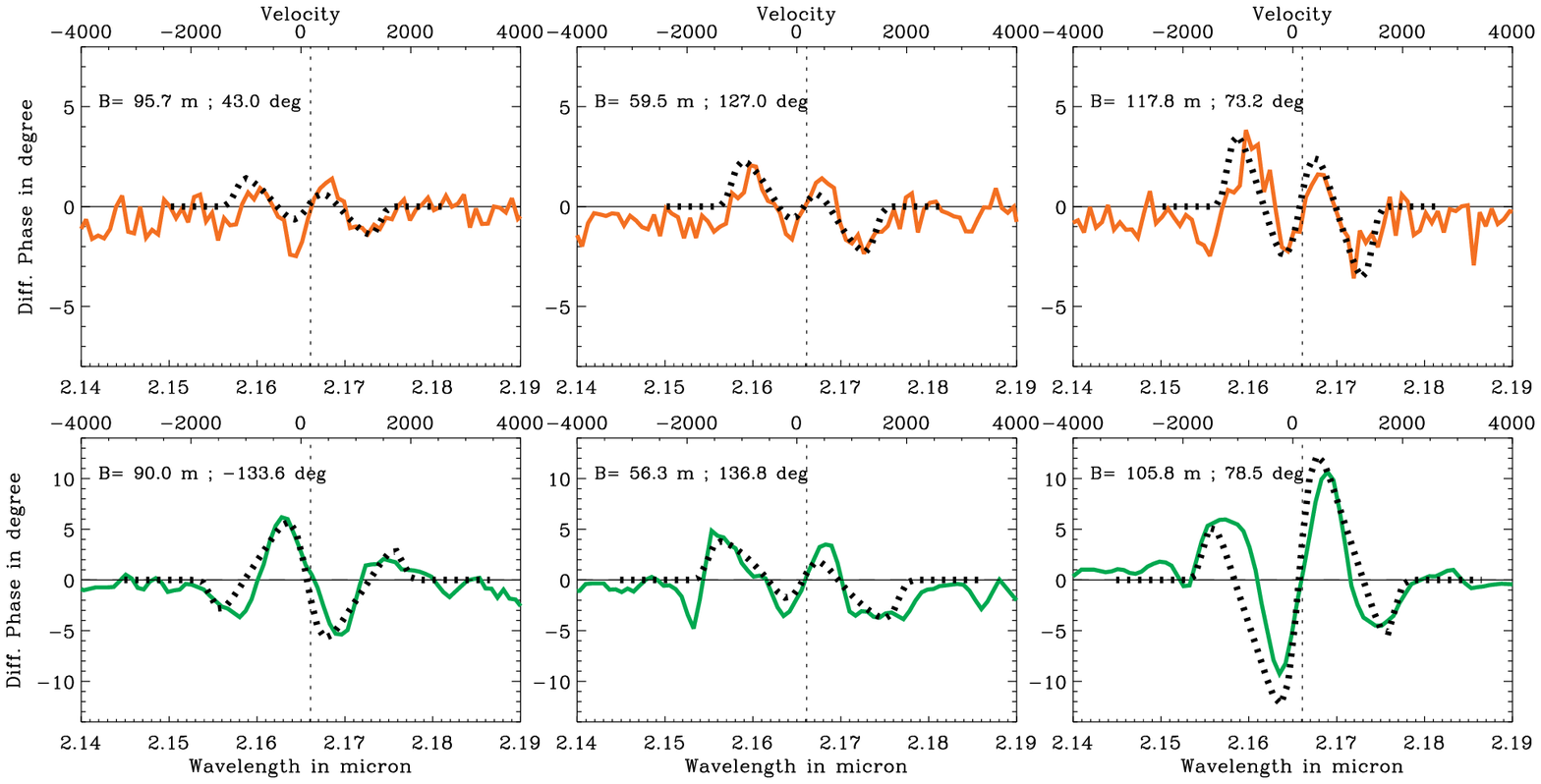}
 \caption[]{{\bf AMBER data.  Top}: comparison between the Br$\gamma$ line at t=8.81 (blue curve), t=28.76d (red curve) and t=35.77d (green curve). {\bf Bottom}: Above - differential visibility comparison between t=28.76d and t=35.77d (scaled to the continuum $V^2$ at t=28.76). Below: same with differential phases. The phases are compared with the phases from the model (dashed line, $\chi_r^2$=1.1 and 1.4, respectively). }
 \label{fig:AMBER}
\end{figure*}

\begin{figure}
 \centering
\includegraphics[width=8.6cm]{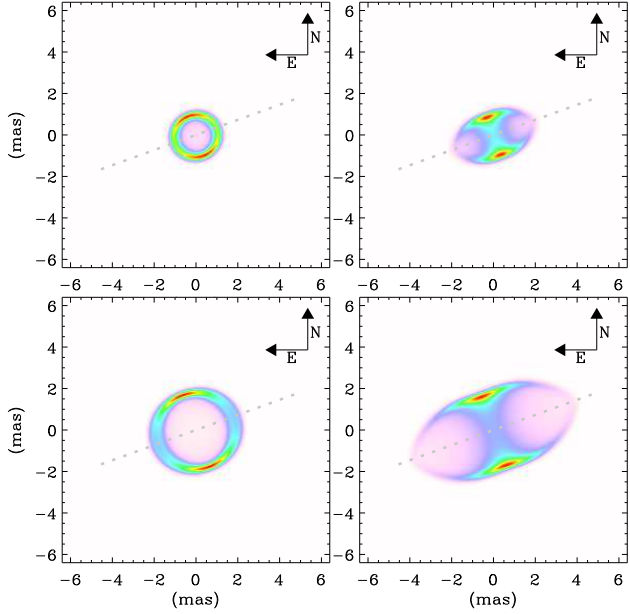}
 \caption[]{ Br$\gamma$ Bipolar-flow model (without central source) seen at i=90\deg\ (right) and i=10\deg\ (left , the best model) and P.A.=110\deg\ (best model).  \label{fig:AMBER2}}
\end{figure}

The heterogeneous data set originating from different instruments provides intriguing data, and is unique in view of the complementary  constraints provided for the analysis. The interpretation of these data can be divided into several key temporal steps.
\begin{itemize}
\item CLASSIC and PIONIER data obtained at t=2.92d, 12.81d and t=13.93d show an extended H and K band source. The fact that \object{T\,Pyx} is resolved so early is intriguing. An hypothesis might be a light echo witnessing a close-by circumbinary environment, adding incoherent flux to the measurement. The expansion rate is then low between the first and second measurements, and a striking difference is observed between the H and the K band inferred diameters. This suggests an advanced decoupling between a shrinking optically thick core and an expanding free-free emitting optically thin envelope. \\
\item The AMBER data obtained at t=28.76d provide evidence that the Br$\gamma$ line forming region projected onto the sky is formed close to the expanding continuum, while being more extended at t=35.77d. The Doppler velocity accelerated from 1300\kms\ to 1600\kms\ in the time interval. A striking differential phase pattern is observed. \\
\item The PIONIER dataset secured at t=48.74d is critical for establishing the two component nature of the emission, consisting in an unresolved stellar source and an extended region whose appearance is circular and symmetric  within error bars.  
\end{itemize}

A face-on bipolar event could account for the ensemble of information described above, in accordance with the finding of a low inclination for the system \citep{2010MNRAS.409..237U} . In particular, the differential phase pattern can be linked to the geometry and kinematics of the ejecta. We developed a `toy' model that provides a good match to the observations. Given two ad-hoc three-dimensional distributions, one for the `emission' of the ejecta and one for the velocity field, we reconstructed intensity maps in narrow spectral bands in the emission line and then computed the corresponding visibilities and differential phases. The intensity map is created considering that the matter was ejected during a brief outburst (best ascribed by a shell), propagating at v$_r$($\theta$). Consequently, the geometry is directly related to kinematics of the ejecta. We used the following radial expansion law:
\begin{equation}
\mathrm{v_r(\theta)}=\mathrm{v_{pole}}+\left(\mathrm{v_{eq}-v_{pole}}\right)\sin{\theta}
\end{equation}
\noindent where $\theta$ is the colatitude, and v$_{pole}$ and v$_{eq}$ are the polar and equatorial radial velocities, respectively.

We also considered an emission decreasing according to a power law of the distance. At a given epoch \textit{t} the 3D intensity distribution is proportional to:

\begin{equation}
{\mathrm I(r,\theta,\phi)}\propto \frac{1}{\mathrm{r^\alpha}}\,\exp{\Bigl[\frac{\mathrm{-(v_r\,t-r)^2}}{\mathrm{2\,\sigma_r^2}}\Bigr]}
\end{equation}

Using this model we were able to fit the Br$\gamma$ differential visibilities and phases, as well as the line profile for the two epochs. The parameters of the best model are: i=15\deg, P.A. of the polar axis of 110\deg, $\alpha$=2 and the velocities for the two epochs:
\begin{itemize}
\item v$_\mathrm{pole}$=1200\,km\,s$^{-1}$ and  v$_\mathrm{eq}$=600\,km\,s$^{-1}$ at t=28.76d
\item v$_\mathrm{pole}$=1600\,km\,s$^{-1}$ and  v$_\mathrm{eq}$=700\,km\,s$^{-1}$ at t=35.77d
\end{itemize}

The fit of the differential phases at the two epochs and the model images are presented in Fig~\ref{fig:AMBER} and the model in Fig~\ref{fig:AMBER2}. The polar and equatorial velocities are in good agreement with the Doppler and sky plane velocities estimated in Sect.\ref{Analysis}. Furthermore, the P.A. of the equatorial plane overdensity is oriented in a direction similar to the P.A. of the faint X-ray nebula \citep{2010MNRAS.404L..26B}.

The face-on bipolar nebula allows one to better understand the curious nebula scrutinized with HST  \citep{2010ApJ...708..381S, 1997AJ....114..258S, 1989ApJ...337..720S}. The knots are concentrated in a ring (3.2-6\arcsec), expanding radially with a velocity in the restricted range of 500-700\kms, and with a mean radial velocity of about 500\kms\citep{1998ApJ...498L..59O}. Deciphering between a projected sphere and a bipolar structure producing a dense, face-on ring is difficult, considering that radial velocity measurements of individual clumps are missing.

Some recent examples suggest that bipolarity in the ejecta of classical/recurrent novae may be relatively frequent: RS\,Oph \citep{2009ApJ...703.1955R,2007ApJ...665L..63B,2007A&A...464..119C}, \object{V445\,Pup} \citep{2009ApJ...706..738W}, \object{V1280 Sco} \citep[Chesneau et al., in prep.]{2008A&A...487..223C} or HR\,Del \citep{2003MNRAS.344.1219H}. A significant difference though exists between the \object{T\,Pyx} and \object{RS\,Oph} environments: the lack of material around \object{T\,Pyx}, witnessed for instance by the lack of hard X-rays \citep{2011ATel.3285....1K}, leads us to favor a bipolarity induced by a process internal to the system, whether by the common envelope interaction with the companion, since the development of the event is relatively slow, or by invoking an intrinsically bipolar ejection related to a spun-up central star \citep{1998MNRAS.296..943P, 1997MNRAS.284..137L}. 

\begin{acknowledgements}
The CHARA Array is funded by the National Science Foundation through NSF grant AST-0908253, by Georgia State University, the W. M. Keck Foundation, the Packard Foundation, and the NASA Exoplanet Science Institute. Research at the  Physical Research Laboratory is funded by the Dept. of Space, Govt. of India. STR acknowledges partial support from NASA grant NNH09AK731.

\end{acknowledgements}

{}

\end{document}